\documentclass[aps,twocolumn,showpacs]{revtex4}
\usepackage{graphicx}
\begin{document}

\title{Influence of the gate leakage current on the stability of organic single-crystal field-effect transistors}
\author{R. W. I. de Boer, N. N. Iosad, A. F. Stassen, T. M. Klapwijk, and A. F. Morpurgo}
\affiliation{Kavli Institute of Nanoscience, Faculty of Applied
Sciences, Delft University of Technology, Lorentzweg 1, 2628 CJ
Delft, the Netherlands}
\date{\today}

\begin{abstract}

We investigate the effect of a small leakage current through the
gate insulator on the stability of organic single-crystal
field-effect transistors (FETs). We find that, irrespective of the
specific organic molecule and dielectric used, leakage current
flowing through the gate insulator results in an irreversible
degradation of the single-crystal FET performance. This
degradation occurs even when the leakage current is several orders
of magnitude smaller than the source-drain current. The
experimental data indicate that a stable operation requires the
leakage current to be smaller than $10^{-9} \ \mathrm{A/cm}^2$.
Our results also suggest that gate leakage currents may determine
the lifetime of thin-film transistors used in applications.

\end{abstract}


\maketitle

The study of organic semiconductor transistors aims at the
development of organic electronics, for its advantages of being
flexible, cheap and suitable for large-area production
\cite{Deleeuw04,Voss00}. So far, considerable research effort has
been focused on the optimization of the organic layer to improve
the performance of thin-film transistors
\cite{Horowitz04,Dimitrakopoulos02,Klauk02}. Much less attention
has been devoted to other important device aspects, such as, for
instance, the choice of the gate insulator.

Recent work has demonstrated that the gate insulator plays an
important role in determining the device performance
\cite{Stassen04,Veres03}. In particular, it has been shown that in
polymer as well as in single-crystal organic transistors the
mobility of charge carriers is systematically larger the lower the
dielectric constant of the gate insulator. This implies that the
use of low-$\epsilon$ dielectrics will result in a higher device
switching speed. In view of this result, it appears useful to
investigate systematically how different properties of the gate
insulator affect the behavior of organic transistors.

In this paper we use organic single-crystal FETs \cite{DeBoer04}
to investigate how a small leakage current through the gate
insulator affects the stability of the device operation.
Specifically, we have investigated the behavior of organic
single-crystal FETs of different molecules (tetracene, rubrene,
perylene) in combination with different dielectrics (Ta$_2$O$_5$,
ZrO$_2$, and SiO$_2$). We find that, irrespective of the specific
molecule and dielectric used, leakage current flowing through the
gate insulator results in an irreversible degradation of the
single-crystal FET operation. The degradation is \textit{not} due
to the electrical breakdown of the insulating layer and it also
occurs when the leakage current is several orders of magnitude
smaller than the source-drain current. From the experimental data,
we conclude that a stable operation of organic single-crystal FETs
requires the current leaking to the FET channel to be smaller than
$10^{-9}$ A/cm$^2$.

The fabrication of the single-crystal FETs used in this work is
based on electrostatic bonding of an organic single-crystal to a
dielectric surface, with pre-fabricated source, drain and gate
contacts. The details are essentially identical to what has been
described in Ref.~\cite{DeBoer03}. Whereas in Ref.~\cite{DeBoer03}
only thermally grown SiO$_2$ was used as gate insulator, here we
have also used sputtered layers of Ta$_2$O$_5$ and ZrO$_2$
deposited in different ways. For both Ta$_2$O$_5$ and ZrO$_2$ we
have investigated FETs in which the dielectric layers were
sputtered from ceramic targets (hereafter referred to as "type I"
oxides; see Ref.~\cite{Nick} for details). For Ta$_2$O$_5$, we
have also investigated the behavior of FETs fabricated on layers
sputtered from a metallic target, in the presence of oxygen in the
sputtering gas, as described in Ref.~\cite{Fleming02} (hereafter
referred to as "type II" Ta$_2$O$_5$). For all FETs discussed here
the sputtered oxide layers were approximately 350 nm thick.

\begin{figure}[b]
\centering
\includegraphics[width=8.5cm]{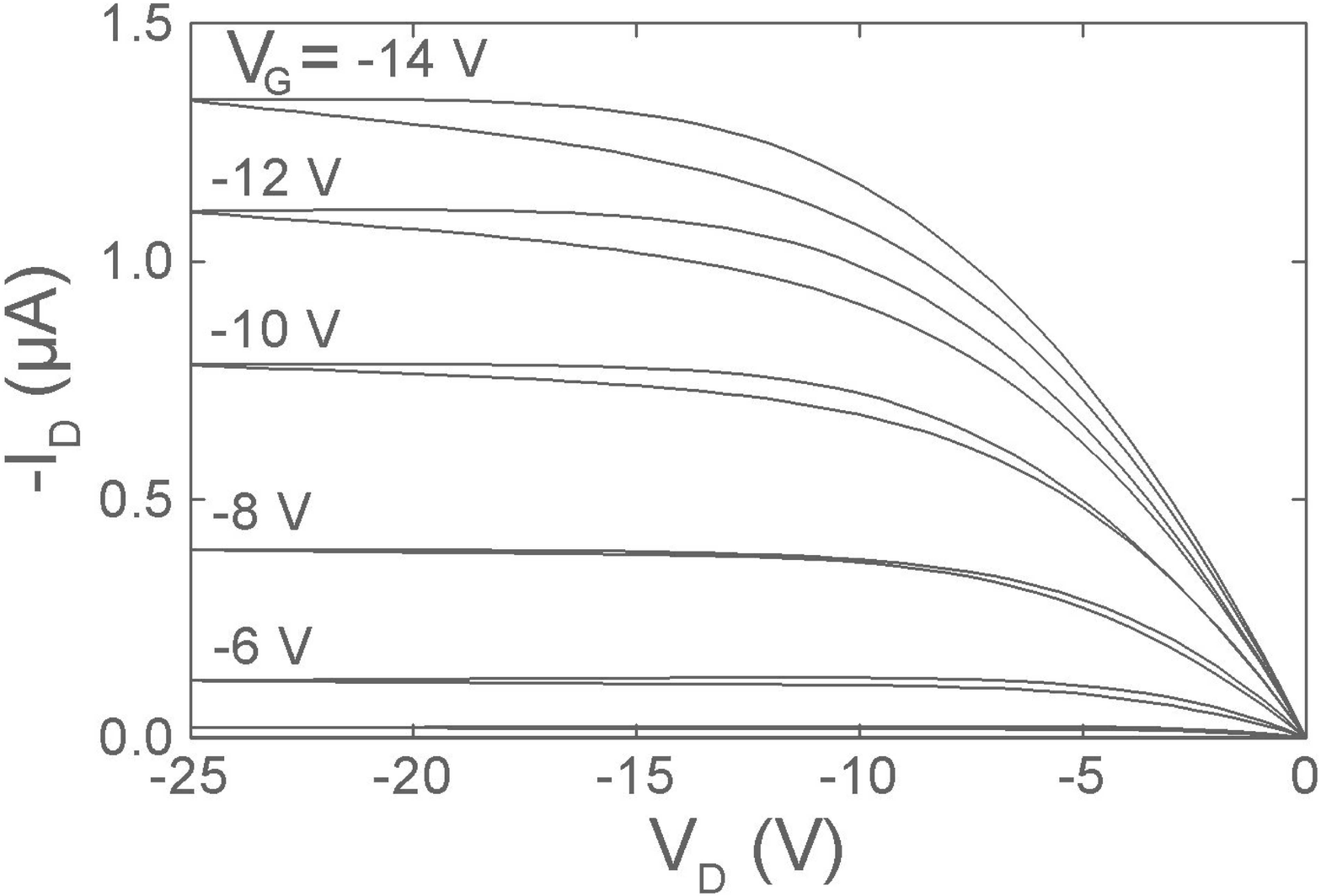}
\caption{Current-voltage characteristics of an organic
single-crystal FET, with tetracene as the organic molecule and
type I Ta$_2$O$_5$ as the gate insulator.
\label{TetraceneTa2O5char}}
\end{figure}

The electrical properties of all the different dielectric layers
were characterized by capacitance and $I$-$V$ measurements (see
\cite{Nick}). From these measurements we obtain a dielectric
constant $\epsilon = 25$ for Ta$_2$O$_5$ (both types) and 23 for
ZrO$_2$, as expected. The breakdown field is comparable for all
layers and typically equal to $E_{\mathrm{bd}} \simeq 4-6$ MV/cm.
The leakage current, on the contrary, is different for the
different layers. Specifically, at a voltage corresponding to
approximately half the breakdown field, the leakage through type I
Ta$_2$O$_5$ is typically in the order of $10^{-6}$ A/cm$^2$,
slightly higher than that through ZrO$_2$, $10^{-7} - 10^{-6}$
A/cm$^2$ and much higher than that flowing through type II
Ta$_2$O$_5$, $< 10^{-9}$ A/cm$^2$ \cite{note1}.

\begin{figure}[t]
\centering
\includegraphics[width=8.5cm]{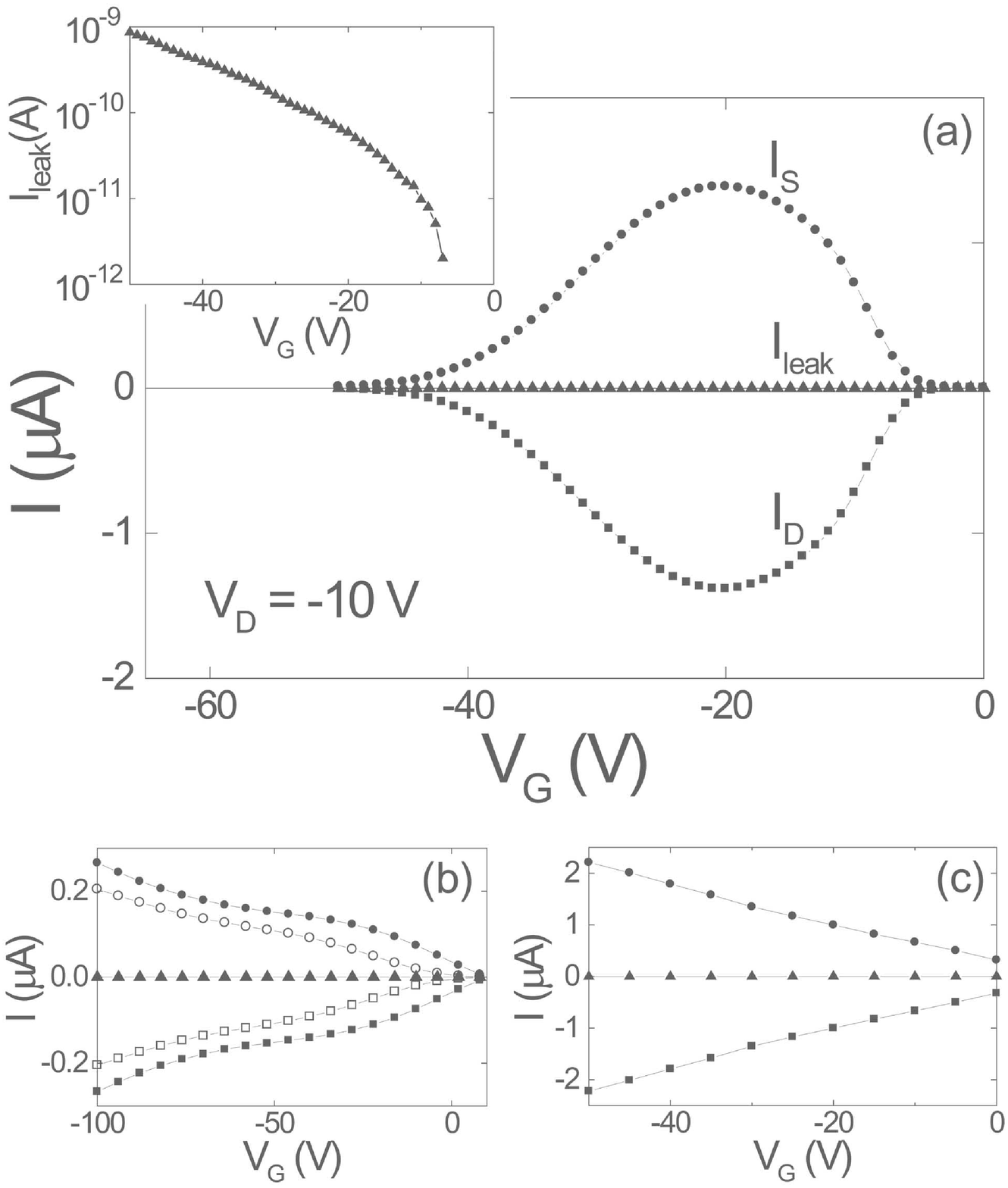}
\caption{(a) Transfer characteristics of a FET on Ta$_2$O$_5$
($V_D = -10$ V). The source and drain current ($I_S$ and $I_D$),
are equal and opposite, since the leakage current
$I_{\mathrm{leak}}$ is orders of magnitude smaller than $I_D$ (see
inset). Device degradation is responsible for the non-monotonic
$I_D$-$V_G$ curve, resulting in the full suppression of $I_D$ at
high $V_G$. (b) Transfer characteristics of a tetracene FET on
ZrO$_2$ ($V_D = -10$ V). The $I_D$-$V_G$ relation is non-linear
and $I_D$ and $I_S$ are lower in the second $V_G$ sweep (open
circles) than in the first sweep (closed circles).
$I_{\mathrm{leak}}$ (triangles) is much smaller than $I_D$ and
$I_S$. Note that the shape of the $I_D$-$V_G$ curves is
characteristic for ZrO$_2$ and differs from that of Ta$_2$O$_5$.
For comparison, (c) shows the transfer characteristics of a
tetracene FET on SiO$_2$ for which $I_S$ and $I_D$ are linearly
related to $V_G$ ($V_D = -10$ V). \label{Tetraceneoxides}}
\end{figure}

Electrical characterization of all devices is performed in the
dark, at room temperature and under vacuum ($10^{-7}$ mbar), using
a HP4156A Semiconductor Parameter Analyzer, in a two-terminal
configuration. Fig.~\ref{TetraceneTa2O5char} shows the transistor
characteristics of a tetracene FET with type I Ta$_2$O$_5$ as gate
dielectric. Upon a superficial inspection, these characteristics
resemble those of tetracene FETs on SiO$_2$ (see
Ref.~\cite{DeBoer03}). The typical mobility values in Ta$_2$O$_5$
devices range from $0.02$ to $0.08$ cm$^2$/Vs, smaller than in
devices on SiO$_2$, as expected in view of the larger dielectric
constant of Ta$_2$O$_5$ \cite{Stassen04}. A first unexpected
difference is however visible in Fig.~\ref{TetraceneTa2O5char}, as
the transistor characteristics exhibit some hysteresis, not
normally observed in high-quality tetracene single-crystal FETs on
SiO$_2$. A much more striking difference between the type I
Ta$_2$O$_5$ and ZrO$_2$ FETs and the SiO$_2$ FETs is clearly
apparent when looking at the $V_G$ dependence of $I_D$ measured at
a fixed $V_D$ value.

Fig.~\ref{Tetraceneoxides}a shows the data obtained from a type I
Ta$_2$O$_5$ FET. Contrary to the usual behavior observed in
tetracene FETs on SiO$_2$ (Fig.~\ref{Tetraceneoxides}c), i.e.
$I_D$ increasing linearly with $V_G$, the source-drain current in
type I Ta$_2$O$_5$ FETs increases, reaches a maximum and then
decreases again. The decrease results in the full suppression of
the source-drain current, unless dielectric breakdown of the
insulator occurs first. This behavior has been observed in all
fabricated devices, irrespective of the organic material used
(tetracene, rubrene, perylene).

For a tetracene FET on ZrO$_2$ $I_D$-$V_G$ curves measured at
fixed $V_D$ are shown in Fig.~\ref{Tetraceneoxides}b. Again, the
$I_D$-$V_G$ relation is markedly non-linear, although full
suppression of the source-drain current is not reached. The shape
of the non-linearity is different from that observed in type I
Ta$_2$O$_5$ devices and is characteristic for our ZrO$_2$
transistors. Also for ZrO$_2$ FETs, the behavior of the
$I_D$-$V_G$ is similar when crystals of different organic
materials are used.

The anomalous behavior reproducibly exhibited by type I
Ta$_2$O$_5$ and ZrO$_2$ FETs originates from \textit{irreversible
device degradation}. Specifically, we observe that, for every
device studied, repeating the measurement of the $I_D$-$V_G$ curve
systematically results in lower measured values of $I_D$ (see
Fig.~\ref{Tetraceneoxides}b). For those type I Ta$_2$O$_5$
transistors in which the increase in $V_G$ is sufficient to fully
suppress the source-drain current (see
Fig.~\ref{Tetraceneoxides}a), no field-effect induced current is
ever observed after the measurement, indicating that the
degradation of the device is complete.

Inspection of the degraded transistors using an optical microscope
does not reveal any visible change in the device. The bulk of the
crystal, the dielectric layer, and the FET circuitry appear to
have all remained intact and the crystal is still well bonded to
the substrate. This suggests that the device degradation is
confined to the first layers of the organic material at the
interface with the dielectric.

\begin{figure}[t]
\centering
\includegraphics[width=8.5cm]{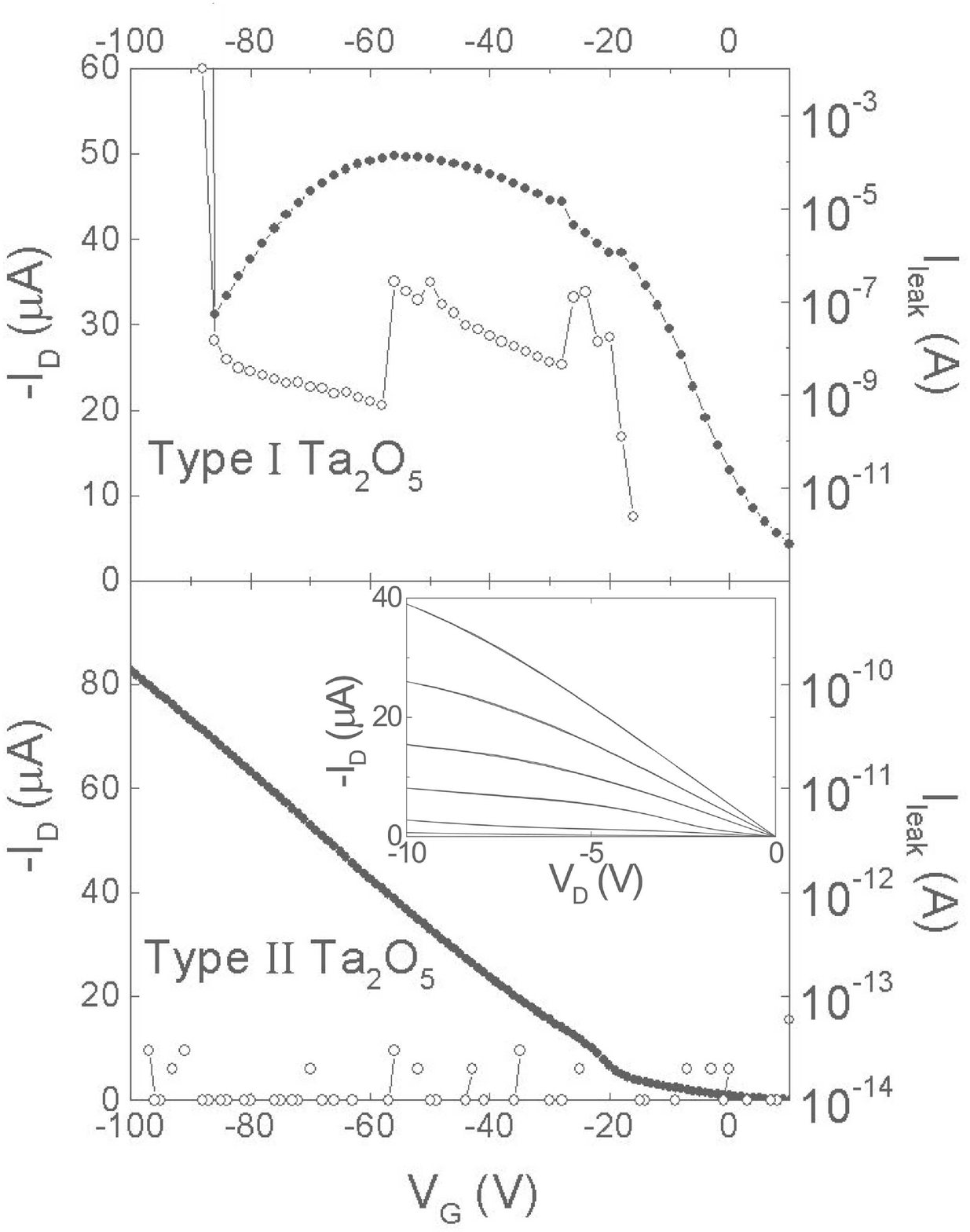}
\caption{$I_D$-$V_G$ curves (closed symbols) of two rubrene FETs
on type I Ta$_2$O$_5$ (upper panel) and type II Ta$_2$O$_5$ (lower
panel) ($V_D = -10$ V in both cases). The open symbols represent
the leakage current. The degradation-induced, non-monotonic
behavior of the $I_D$-$V_G$ curve observed for the type I
Ta$_2$O$_5$ FET is absent in the type II Ta$_2$O$_5$ device. Note
that the leakage current in the two devices differs by orders of
magnitude, despite the comparable crystal surface area ($\sim
\mathrm{mm}^2$) and Ta$_2$O$_5$ thickness ($\sim 350$ nm). The
inset shows that FETs on type II Ta$_2$O$_5$ exhibit
hysteresis-free electrical characteristics.
\label{RubreneTa2O5sweep}}
\end{figure}

To determine the cause of device degradation it is revealing to
compare the behavior of single-crystal FETs fabricated using type
I and type II Ta$_2$O$_5$. Fig.~\ref{RubreneTa2O5sweep} shows the
results of $I_D$-$V_G$ sweeps for two rubrene single-crystal FETs
fabricated using the two different oxides. Similar to what we have
shown for the tetracene FETs in Fig.~\ref{Tetraceneoxides}, the
FET on type I oxide exhibits a non-linear, non-monotonic
$I_D$-$V_G$ relation up to the dielectric breakdown voltage (at
$V_G \approx -80$ V for this sample). Conversely, for the rubrene
FET on type II Ta$_2$O$_5$, $I_D$ scales linearly with $V_G$ in a
large range of values above the threshold voltage, up to the
maximum voltage reached in the experiment ($V_G = -100$ V,
corresponding to a charge density of $\simeq 5 \times 10^{13}$
holes/cm$^2$). For this FET, multiple measurements of the
$I_D$-$V_G$ curve reproducibly give the same result. Note also, in
the inset of Fig.~\ref{RubreneTa2O5sweep}, that the transfer
characteristics of the rubrene FET on type II Ta$_2$O$_5$ are
fully hysteresis-free, as is also typical for high-quality
transistors fabricated on SiO$_2$. In short, contrary to what
happens to devices based on type I Ta$_2$O$_5$, for FETs
fabricated on type II Ta$_2$O$_5$ device degradation does not
occur. Since the main difference between type I and type II
Ta$_2$O$_5$ layers is the much higher level of leakage current
observed in the type I layers, this observation suggests that the
current leaking through the gate insulator is the cause for the
device degradation.

To further investigate the origin of the FET degradation, we have
also studied FETs fabricated on bilayers consisting of a 350 nm
thick layer of type I Ta$_2$O$_5$ (ZrO$_2$) covered with a 25 nm
thin top layer of ZrO$_2$ (type I Ta$_2$O$_5$), so that the
organic crystal is in contact with the thin top layer. For these
FETs, the shape of the $I_D$-$V_G$ curve is similar to that
observed in FETs where the thin top layer is not present. These
experiments indicate that the details of the device degradation
are determined by the thick oxide layer and not by the material
directly in contact with the organic crystals. This observation
rules out the possibility that a chemical reaction between
molecules and dielectric material is causing the device
degradation and confirms the role of the leakage current, since in
these oxide bi-layers it is the thick layer that determines the
magnitude of $I_{\mathrm{leak}}$.

We conclude that damage to the organic crystal induced by current
leaking through the gate insulator is the cause for the device
degradation \cite{note2}. This conclusion is further supported by
the absence of degradation in single-crystal FETs fabricated on
SiO$_2$, in which the leakage current is undetectably small. It is
also consistent with the larger degradation observed in type I
Ta$_2$O$_5$ FETs as compared to ZrO$_2$ devices, since the leakage
current through ZrO$_2$ is typically almost an order of magnitude
less than in type I Ta$_2$O$_5$.

It is worth noting that degradation occurs even when \textit{the
leakage current is several orders of magnitude lower than the
source-drain current}. Specifically, our data quantitatively show
that in organic single-crystal FETs gate leakage currents larger
than approximately $10^{-9} \ \mathrm{A/cm}^2$ \cite{note3}
systematically result in irreversible device degradation. This
conclusion poses a clear constraint on the design of properly
functioning single-crystal FETs. It is possibly also relevant for
organic thin-film transistors, as it suggests that the gate
leakage current is an important factor in determining the device
lifetime.

In conclusion, we have shown that leakage current from the gate
electrode causes irreversible degradation of organic
single-crystal FETs, even when it is orders of magnitude smaller
than the source-drain current. This poses a clear constraint for
the design of single-crystal transistors currently used to
investigate the intrinsic electronic properties of organic
semiconductors. As a by-product of this work, we have successfully
fabricated single-crystal devices operating up to a charge density
of at least $5 \times 10^{13} \ \mathrm{carriers/cm}^2$ ($\sim 1$
carrier per 10 molecules), which will enable the investigation of
organic single-crystal FETs at high carrier density.

We acknowledge FOM for financial support. The work of AFM is part
of the NWO Vernieuwingsimpuls 2000 program.


\begin{thebibliography}{99}

\bibitem{Deleeuw04} G. H. Gelinck, T. C. T. Geuns, and D. M. de Leeuw, Nature Mater. \textbf{3}, 106 (2004)
\bibitem{Voss00} D. Voss, Nature \textbf{407}, 442 (2000)
\bibitem{Horowitz04} G. Horowitz, J. Mater. Res. \textbf{19}, 1946 (2004)
\bibitem{Dimitrakopoulos02} C. D. Dimitrakopoulos, and P. R. L. Malenfant, Adv. Mater. \textbf{14}, 99 (2002)
\bibitem{Klauk02} H. Klauk, M. Halik, U. Zschieschang, G. Schmid, W. Radlik, and W. Weber, J. Appl. Phys. \textbf{92}, 5259 (2002)
\bibitem{Stassen04} A. F. Stassen, R. W. I. de Boer, N. N. Iosad, and A. F. Morpurgo, cond-mat/0407293.
\bibitem{Veres03} J. Veres, S. D. Ogier, S. W. Leeming, D. C. Cupertino, and S. M. Khaffaf, Adv. Funct. Mater. \textbf{13}, 199 (2003)
\bibitem{DeBoer04} R. W. I. de Boer, M. E. Gershenson, A. F. Morpurgo, and V. Podzorov, Phys. Stat. Sol. \textbf{201}, 1302 (2004)
\bibitem{DeBoer03} R. W. I. de Boer, T. M. Klapwijk, and A. F. Morpurgo, Appl. Phys. Lett. \textbf{83}, 4345 (2003)
\bibitem{Nick} N. N. Iosad, G. J. Ruis, E. V. Morks, A. F. Morpurgo, N. M. van der Pers, P. F. A. Alkemade, and V. G. M. Sivel, J. Appl. Phys. \textbf{95}, 8087 (2004)
\bibitem{Fleming02} K. Chu, J. P. Chang, M. L. Steigerwald, R. M. Fleming, R. L. Opila, D. V. Lang, R. B. Van Dover, and C. D. W. Jones, J. Appl. Phys. \textbf{91}, 308 (2002)
\bibitem{note1} The microscopic origin of the difference in leakage
current between Type I and Type II Ta$_2$O$_5$ is probably due to the larger
deposition rate that can be achieved in sputtering from a metal
target, which results in the inclusion of less impurities in the
sputtered layers.
\bibitem{note2} The specific microscopic process responsible for the
degradation of the organic material remains to be understood. One
possible mechanism is that high-energy electrons leaking through
the gate insulator physically break individual molecules at the
crystal surface, thus causing the appearance of a very large
number of traps in the FET active regions.
\bibitem{note3} A more precise quantification is difficult because an
unknown fraction of leakage current flows directly to the source
or drain contacts and does not cause damage to the organic
crystal.

\end{thebibliography}
\end{document}